\documentclass[review]{elsarticle}

\usepackage{lineno,hyperref}
\usepackage{graphicx}
\usepackage{epstopdf}
\modulolinenumbers[5]

\journal{Journal of \LaTeX\ Templates}

\begin{document}

\begin{frontmatter}

\title{Euro-VO - Coordination of Virtual Observatory activities in Europe \tnoteref{mytitlenote}}

\author{Fran\c coise Genova}
\ead{francoise.genova@astro.unistra.fr}
\author{Mark G. Allen}
\ead{mark.allen@astro.unistra.fr}
\address{Observatoire Astronomique de Strasbourg, Universit\'e de Strasbourg - CNRS UMR 7550, 11 rue de l'Universit\'e, Strasbourg, France}

\author{Christophe Arviset}
\ead{christophe.arviset@esa.int}
\address{ESA-ESAC, POBOX 78, 28691 Villanueva de la Canada, Madrid, Spain}

\author{Andy Lawrence}
\ead{al@roe.ac.uk}
\address{Institute for Astronomy, University of Edinburgh, Scottish Universities Physics Alliance (SUPA), Royal Observatory, Blackford Hill, Edinburgh EH9 3HJ, U.K.}

\author{Fabio Pasian}
\ead{pasian@oats.inaf.it}
\address{INAF - Astronomical Observatory of Trieste, Via G.B. Tiepolo 11, 3114, Trieste, Italy}

\author{Enrique Solano}
\ead{esm@cab.inta-csic.es}
\address{Centro de Astrobiolog\'{\i}a (INTA-CSIC). Departamento de Astrof\'{\i}sica. Campus Villafranca. P.O. Box 78, E-28691 Villanueva de la Ca\~ nada, Madrid, Spain; Spanish Virtual Observatory}

\author{Joachim Wambsganss}
\ead{jkw@ari.uni-heidelberg.de}
\address{Zentrum f\"{u}r Astronomie der Universit\"{a}t Heidelberg (ZAH), Astronomisches Rechen-Institut (ARI), M\"{o}nchhofstr. 12-14, 69120 Heidelberg, Germany}



\cortext[mycorrespondingauthor]{Fran\c coise Genova}


\begin{abstract}
The European Virtual Observatory Euro-VO has been coordinating European VO activities through a series of projects co-funded by the European Commission over the last 15 years. The bulk of VO work in Europe is ensured by the national VO initiatives and those of intergovernmental agencies. VO activities at the European level coordinate the work in support of the three ``pillars'' of the Virtual Observatory: support to the scientific community, take-up by the data providers, and technological activities. Several Euro-VO projects have also provided direct support to selected developments and prototyping. This paper explains the methodology used by Euro-VO over the years. It summarizes the activities which were performed and their evolutions at different stages of the development of the VO, explains the Euro-VO role with respect to the international and national levels of VO activities, details the lessons learnt for best practices for the coordination of the VO building blocks, and the liaison with other European initiatives, documenting the added-value of European coordination. Finally, the current status and next steps of Euro-VO are briefly addressed.

\end{abstract}

\begin{keyword}
virtual observatory tools, standards, astronomical databases: miscellaneous
\MSC[2010] 00-01\sep  99-00
\end{keyword}

\end{frontmatter}


\section{Introduction: Why a European Virtual Observatory project?}

The digital revolution is bringing a change in paradigm in the way science is done. Astronomy has long been at the forefront for widespread sharing and re-use of data, with a new step towards interoperability and data integration through the development of the astronomical Virtual Observatory (VO). The concept of a Virtual Observatory is that astronomical data worldwide should be accessible through common protocols, with interoperability standards for metadata and tools so that access and use of the data become seamless. VO is embraced as a world-wide community-based initiative that is transforming and restructuring the way astronomy research is done.

The VO has now reached a significant level of maturity and provides an operational framework for interoperable and efficient access to world-wide astronomical data and services. The construction of the VO has followed a predominantly bottom-up, incremental, pragmatic approach that has been carefully matched to the needs of the astronomical community. This has been supported by national VO initiatives and data centres, with a significant contribution from the European VO initiatives. Having reached this level of maturity, it is now timely to write its history and to describe how it was built and the lessons learnt over the years. This paper is focussed on the European Virtual Observatory, Euro-VO. The main role of Euro-VO has been to coordinate European VO activities, with also for some of the European projects support to technological activities. This paper will highlight the main VO building blocks and lessons learnt on how to coordinate them.

The astronomical VO was conceived from the beginning as a global endeavour, and European participants were among the founding members together with similarly interested US-based participants. The two colloquia which set the scene in 2000, \emph{Virtual Observatories of the Future} and \emph{Mining the Sky}, were organised respectively in Pasadena (USA) on June 13th-16th and in Garching near Munich (Germany) on July 31st-August 4th, both with an international Programme Committee. Then the International Virtual Observatory Alliance (IVOA) was founded during the Conference \emph{Towards an International Virtual Observatory} held in 2002 in Garching, June 10th-14th.

Formal participation in the VO is based on national and international collaborations gathering development teams and data providers, as stated in the ``Guidelines for participation of the International Virtual Observatory Alliance'' (Hanisch et al. 2010 \citep{IVOAParticipation}). VO initiatives from several European countries are IVOA members, as is the European Space Agency (ESA), which is an intergovernmental organisation. The potential usefulness of the coordination of VO activities at the European level was recognised very early. A mechanism was needed to bring the relevant parties together.

Funding by the European Commission through its successive Framework Programmes (FP) has a strong structuring role for astronomy in the European countries, as well as the two intergovernmental organisations ESA and ESO which pool resources from their member countries to realize large space- and ground based programmes. It thus appeared natural to set up a coordination of the VO activities at that level, which soon became the European Virtual Observatory Euro-VO. Inclusive federation of complementary expertise with a light organisational structure and an evolving partnership has been the foundation of the Euro-VO. As a starting point, the two intergovernmental organisations ESA and ESO brought in experience on how to deal with observational data from large ground and space-based telescopes, Strasbourg astronomical data centre CDS had anticipated several aspects of the Virtual Observatory (for instance Aladin already provided access to distant observatory archives, and employed a registry of resources (Fernique, Ochsenbein and Wenger 1998 \citep{1998ASPC..145..466F})), and the UK VO project AstroGrid had just started. These were the main partners of the first European project. Other VO initiatives progressively started in other countries, and Italy, Spain and Germany (respectively the Italian Virtual Observatory VObs.it, the Spanish Virtual Observatory SVO and the German Astrophysical Observatory GAVO) joined Euro-VO.

The  \emph{Astrophysical Virtual Observatory} (AVO) project was successfully submitted for funding to the European Commission in 2001 (Quinn et al. 2002 \citep{2002SPIE.4846....1Q}). This allowed a close collaboration between the national initiatives, ESA, and ESO. As expected, the European initiatives brought their complementary expertise, and coordination and collaboration allowed them to reach a critical mass for the different aspects of the activities. We found that activities developed in collaboration at the European level worked well as templates for activities at national level, and technical collaborations between teams from different countries strengthened the impact of the work of the individual teams.

It was soon realized that the VO aim of providing seamless access to the wealth of astronomical resources, was well aligned with key objectives of the European Union. One of the main impacts of the VO is to offer an equal opportunity to access the best data and tools to scientists from all countries, and in particular to astronomers of all European countries, including peripheral and outermost regions. All data and service providers, large agencies, and small teams willing to share their knowledge, can contribute. The VO is thus a powerful vehicle for the integration of the national European astronomical communities into a single community, which is actually world-wide.

The European project also gave good visibility to VO development across Europe. This has been very important to get the VO taken into account in the European strategy for Astronomy \citep{Roadmap2008} established in 2008 by the Astronet ERA-NET\footnote{ASTRONET was created in 2005 by a group of European funding agencies in order to establish a strategic planning mechanism for all of European astronomy. As of 2014, the ASTRONET consortium has eleven participants, twenty-one Associates and two Forum members.}, which gathers the astronomy funding agencies.

The paper is organised as follows: Section 2 briefly describes the suite of European projects which supported the development of Euro-VO over the years, and the evolution of the activities while the project was maturing; Section 3 explains the specific role of Euro-VO in the global VO landscape, and its positioning with respect to the IVOA and the national initiatives; Section 4 details the lessons learnt from the building of Euro-VO, and identify best practices for coordinating VO activities; Section 5 explains the Euro-VO role in a wider European context, its liaison with ``nearby'' disciplines and projects which work on generic building blocks of the data infrastructure. Section 6 concludes by discussing the current status of Euro-VO and the next steps.

\section{The European VO projects: building the European contribution to the Virtual Observatory}

The VO is one of the research infrastructures of astronomy, but it was a novel concept when it emerged and there was no pre-existing model to follow to develop it. The European projects progressively went through different phases, from R\&D and proof of concept to operations, and defined the activities required to develop and maintain the European VO. The projects concerned are briefly described in this section, to show how the current Euro-VO strands of work progressively took shape. More details on how the activities were performed will be given in Section 4 which explains the lessons learnt.

\subsection{The early stages of Interoperability}
The first key contribution of European funded projects to the construction of the VO was through the \emph{Interoperability Working Group} set up by the OPTICON Thematic Network\footnote{Fifth Framework Programme, HPRI-CT-1999-4002} in 2001, that was led by CDS. The global nature of the VO was understood from the start, and this Working Group included participants from the USA and Canada. It was the first international forum to discuss astronomical interoperability standards in the VO context.

A meeting held in Strasbourg in 2002, January 28th-29th, was the first exchange of views between the emerging VO projects which were just beginning to be funded (Europe, UK and USA). The meeting was a starting point for Europe-wide collaboration in the development of the VO, with representatives from France, Germany, Italy, the Netherlands, Spain, UK, ESA and ESO, going beyond the initial European VO partnership set up a few months before in the first VO project AVO, which will be described below. Several of the participants of this initial meeting have since played a key role in VO initiatives that were set up later in their country or Agency.

The Strasbourg meeting allowed a detailed discussion of a standard format for tabular data that was initially proposed in collaboration between the CDS and US-NVO. This was the starting point for VOTable (Ochsenbein et al, 2004 \citep{2004tivo.conf..118O}), an XML-based standard for the exchange of tabular data, which became the first VO standard in March 2002. VOTable was derived from the \emph{astrores} XML format (Ochsenbein et al., 2000 \citep{2000ASPC..216...83O}), itself modeled on the FITS table format (Cotton et al., 1995 \citep{1995A&AS..113..159C})\footnote{VOTable overcomes limitations of FITS binary table usage in a distributed-computing environment, in particular the dataset size is not required in the header, which is an issue for remote data streams. It also allows expression of passwords or other identity information. VOTable can be used either to encapsulate FITS binary tables or to re-encode the metadata.}. VOTable was taken up by the IVOA after it was created and is currently at Version 1.3. It has been (with the pre-existing FITS standard) a key asset for the development of the VO, allowing data exchange and thus a first level of interoperability from the early stages, and it has been massively used since then.

The OPTICON Interoperability Working Group was superseded by the IVOA when it was created a few months later in June 2002. The WG had demonstrated again what had been already experienced in other contexts such as the World Wide Web Consortium W3C or the Internet Engineering Task Force IETF, and also in astronomy with FITS: the usefulness and efficiency of international discussions on standards, when the work is well prepared by the involvement of knowledgeable people and submitted to wide, open discussion by the community - the way the IVOA has been working since its creation.

\subsection{The Euro-VO projects}

European funding is organised in cycles, the so-called Framework Programmes of the European Commission (EC). Euro-VO\footnote{\url{http://www.euro-vo.org}} was supported by several framework programmes, first in the Research, then in the e-Infrastructure programmes. Figure 1 displays the timeline of the successive Euro-VO projects funded by the EC, together with the EC Framework Programmes and the structuring European activities Astronet and OPTICON, also cited in the paper. Table 1 provides information about the EC-funded projects including dates and partners.

\begin{figure*}
\begin{center}
\rotatebox{90}{\includegraphics[height=15cm]{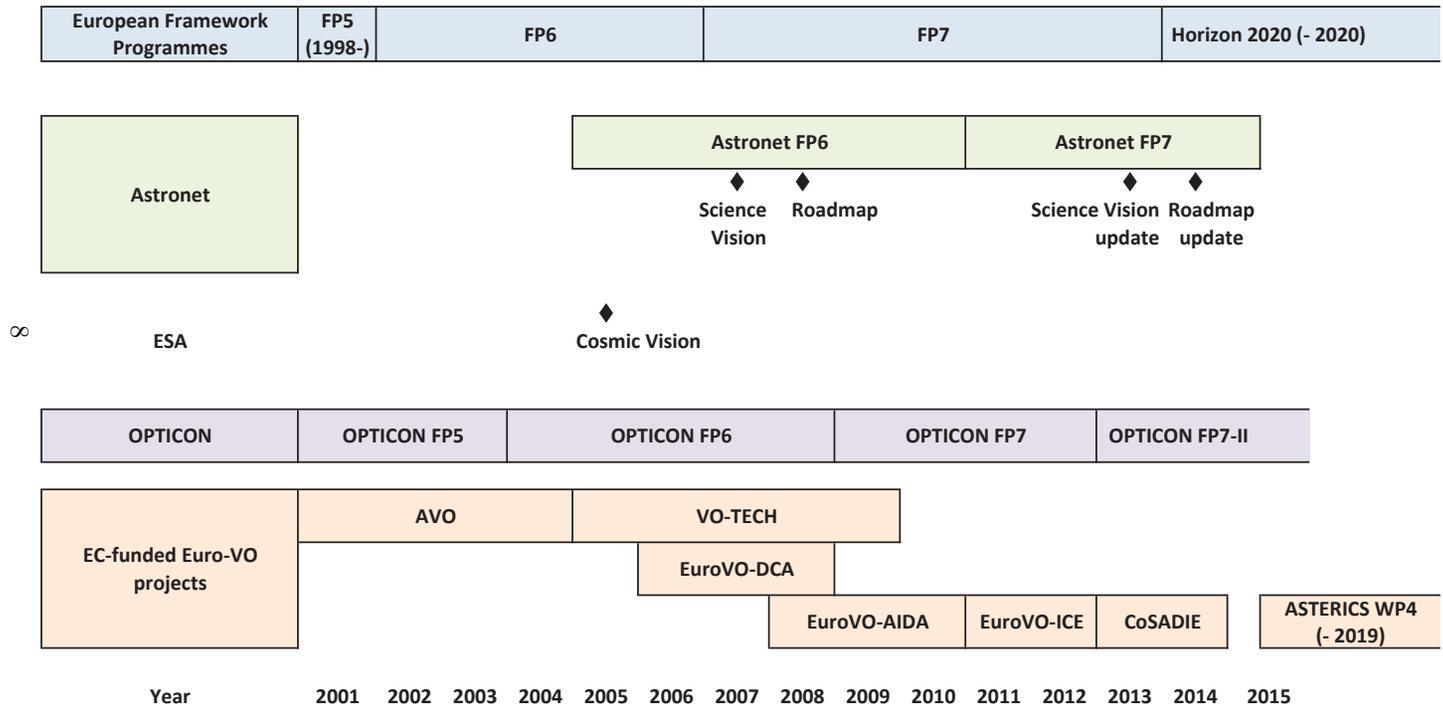}}
\end{center}
\caption{Timeline of the Euro-VO projects, related European structuring activities and strategy documents (time is rounded)}
\end{figure*}

\begin{table}
\centering
\rotatebox{90}{
\begin{tabular}{c|p{1cm}|p{2cm}|p{2.5cm}|l}
Project & FP & ID & Start-End Date & Partners \\ \hline
AVO & FP5 & HPRI-CT- & 2001-2004 & ESO \\
 & & 2001-50030 &  & ESA (ST-ECF), AstroGrid, Univ. Louis Pasteur/CNRS (CDS) \\
 & & & & Jodrell Bank, Terapix  \\ \hline
VO-TECH & FP6 & 011892 & 2005-2009 & Univ. Edinburgh \\
 & & & & Univ. Cambridge, Univ. Leicester, CNRS/ULP (CDS), ESO, INAF \\ \hline
EuroVO-DCA & FP6 & RI031675 & 2006-2008 & CNRS (CDS) \\
 & & & & ESA, ESO, INAF (VObs.it), INTA (SVO), Univ. Leicester (AstroGrid), \\
 & & & & Univ. Groningen, MPG (GAVO) \\ \hline
EuroVO-AIDA & FP7 & 212104 & 2008-2010 & CNRS (CDS) \\
 & & & & ESA, ESO, INAF, INTA, Univ. Edinburgh (AstroGrid), \\
 & & & & Univ. Groningen, Univ. Heidelberg (GAVO) \\ \hline
EuroVO-ICE & FP7 & 261541 & 2010-2012 & CNRS (CDS) \\
 & & & & ESO, INAF, Univ. Edinburgh (AstroGrid) \\ \hline
CoSADIE & FP7 & 312559 & 2012-2015 & CNRS/U. Strasbourg (CDS)\\
 & & &  & INAF, INTA, Univ.Edinburgh (AstroGrid), Univ. Heidelberg (GAVO) \\
\end{tabular}
}
\caption{EC-funded Euro-VO projects. The coordinating partner appears first in the list.}
\end{table}

\subsubsection{The early phases of Euro-VO}

The early phases of Euro-VO are described in Genova (2009) \cite{2009MmSAI..80..548G}. The first European project fully devoted to the Euro-VO, \emph{Astrophysical Virtual Observatory} (2002-2004), set up a Science Working Group and produced a Science Reference Mission. AVO provided proofs of concept of the VO via successful yearly science demonstrations, with objectives defined by the Science Working Group. The demonstrations were based on prototype standards, portals and tools, with the first usage of Aladin as VO portal for images. The last science demonstration led to the publication of the first refereed paper relying on an end-to-end usage of the VO (Padovani et al., 2004 \citep{2004A&A...424..545P}).

AVO led to the understanding that the coordination of VO activities relied on a ``triangle'': support for scientific users, support for data providers, and technological activities. This structured the projects in the following phases (Padovani, 2006 \citep{2006ASPC..351..771P}).

The AVO was succeeded by two projects of the Sixth Framework Programme, dealing with two of the required strands of work of the VO construction, in addition to the communication with scientific users and the gathering of science requirements which had been explored by the AVO:

 \begin{itemize}

 \item
Technological development, with the VOTECH Design Study\footnote{Sixth Framework Programme, \emph{Support for Research Infrastructures, Design Studies}. } (2005-2009), which assessed technological aspects of the construction of Euro-VO with four domains of study: ``Architecture'', ``New user tools'', ``Intelligent resource discovery'', ``Data exploration'', together with the technology projects of the Euro-VO partners. VOTECH defined a reference architecture for the VO, identifying the architecture components, tools, and relevant technologies and standards, and studied some of these components. It also began to coordinate European technical input in IVOA. The project worked in close collaboration with ESA.
 \item
Engagement with data centres, with the \emph{Euro-VO Data Centre Alliance} Collaboration Action\footnote{Sixth Framework Programme, \emph{eInfrastructures}, Call \emph{Communication Network Development}.} (EuroVO-DCA, 2006-2008). EuroVO-DCA aimed at coordinating the first integration of European data centres in the VO framework, and also tackled for the first time the inclusion of modelling results in the VO and the connection of data centres with a distributed computational e-infrastructure.  Data providers are a key component of the VO since they populate it with data. They have to be informed about the developments and supported in their uptake, and it is critical to gather their feedback and requirements since they know the characteristics of their data holdings and their users' needs, and they are concerned about the level of effort required to publish their data in the VO framework. EuroVO-DCA performed the first census of European data centres. It also included explicitly for the first time  liaisons with the astronomical community in European countries beyond the project partners, which has been a standing item of all the projects since then. Collaborations and VO Days organised with the Czech Republic, Lithuania, Poland, etc served to raise awareness and in some cases to create a national proto-VO.

\end{itemize}

 \subsubsection{Transition to operations}

The Euro-VO transition to operations has been supported by three projects under the Seventh Framework Programme. The first one was the Integrated Infrastructure Initiative (I3) \emph{Euro-VO Astronomical Infrastructure for Data Access}\footnote{Call INFRA-2007-1.2.1 Call \emph{Scientific Digital Repositories}.}  (EuroVO-AIDA, 2008-2010). The activities were organised, as required in I3 projects, into ``networking'', ``service'' and ``joint research'' (i.e. Research \& Technology) activities. For the first time, EuroVO-AIDA integrated all the activities necessary to develop the Euro-VO identified by the previous projects, with different strands of work to deal with VO users and the VO community itself. Two kinds of users had been identified as explained above: the astronomers who use VO-enabled data and tools in their everyday research, and the data providers who use the VO framework to publish their data holdings. The VO community are the individuals, teams and initiatives which produce VO standards and VO-enabled tools. The high-level objectives of EuroVO-AIDA were thus:

 \begin{itemize}
 \item
Construction of a community of science users, and the gathering of their needs

\item
Large scale deployment of IVOA protocols and standards by the astronomical data repositories

\item
Maintenance and development of the VO technical infrastructure, including VO standards and tools, and assessment of the usage of new technologies

\end{itemize}

Dissemination of information about the VO to scientists and data providers (schools, seminars, VODays, etc) was tackled through a combination of ``networking'' and ``service'' activities. EuroVO-AIDA also included European participation in the IVOA, and service activities such as the development and maintenance of a Euro-VO registry of resources, harvesting and harvested by the other VO registries.

The project included for the first time activities aimed at higher education and the general public, for which it was chosen to adapt tools used by professional astronomers, to give a feeling as close as possible to research work. The underlying aim is that the appeal of astronomy for the general public makes it a good vehicle to catch the attention of youths and possibly attract them to scientific and technical carriers - an important societal challenge. Tutorials were developed, allowing the solution of scientific questions using the VO, and extensively tested in classrooms, attracting indeed lots of attention from the students.

EuroVO-AIDA also explored possible liaison with other disciplines, building on contacts established within the VO initiatives of several partner countries (Section 5).

The two next projects, \emph{Euro-VO International Cooperation Empowerment}\footnote{Seventh Framework Programme, Programme \emph{Infrastructures, supporting measures}.}  (EuroVO-ICE, 2010-2012) and \emph{Collaborative and Sustainable Astronomical Data Infrastructure for Europe}\footnote{Seventh Framework Programme,  Programme \emph{Coordination actions, conferences and studies supporting policy development, including international cooperation, for e-Infrastructures}.} (CoSADIE, 2012-2015) were small Coordination Actions which allowed Euro-VO to maintain the coordination of European VO activities and the representation of Europe in the IVOA, and to explore paths towards medium term sustainability. CoSADIE concentrated on a number of specific areas, namely, ``Increasing awareness and gathering requirements from the users and data producers'', ``Coordinating technical activities and defining the technical needs to maintain the VO framework'' and ``Outreach towards education and the general public interested in astronomy''. These areas cover the different aspects of the VO-related work and they gather elements to feed the sustainability assessment.

The following sections summarize the findings of the projects on the role of Euro-VO in the VO landscape and lessons learnt for how to most effectively perform the different kinds of VO-related activities.

\section{The role of Euro-VO in the VO landscape}

Euro-VO is an element of the VO landscape which comes between the international layer, the IVOA, and the national VO initiatives. The initial reasons to set up a European VO project were explained in Section 1: the structuring role of European funding and of the European intergovernmental agencies. The impact of Euro-VO activities along the years shows that there is need for an additional ``continental'' layer in the VO landscape, in Europe at least. The initial choice to create a light, inclusive structure for Euro-VO has proven to be an efficient one. It allowed us to pool expertise and interface easily with national initiatives (whatever their organisational structure is), and with the intergovernmental agencies. The light and inclusive nature of the structure allows an evolving partnership, and also reduced the organising overhead at the European level to a minimum.

The \textbf{international level} is of course of vital importance, as the VO will only work with uniform agreement on standards and protocols. Co-operation and agreement on interoperability standards is managed through the International Virtual Observatory Alliance.

At the \textbf{national level}, the situation is different in each of the 28 member states of the European Union.  In the rare countries with no astronomical scientific community in the discipline, VO-enabled teaching in schools maintains interest in astronomy and physics courses, and can have an impact on increasing awareness about and appeal of technology and technological careers. Nearly all countries in the European Union have an astronomical community, and potential users of the VO for its scientific research. Moreover, observatories which operate telescopes or instruments, or teams who want to share their results, are also interested in publishing data and services in the VO. Grouping of efforts at the European level means that many countries do not feel the need to support a national VO initiative, as they may rely on Euro-VO to support uptake by their scientists and data providers.

The national VO initiatives are where the bulk of the work takes place (here we include projects run by the inter-governmental agency, ESA). These initiatives are all different and their aims and organisation depend strongly on their national/agency priorities, and on the way their national research and funding schemes are organised. The VO initiatives are the key players in the development of the VO standards and tools, and national or Agency funding must provide the core support for the data centres which populate the VO with data and services. In addition, the national initiatives are close to their national science communities and so are best placed to organise large-scale support for their users and data centres.

The \textbf{European level} binds together the national efforts and increases their effectiveness. Euro-VO\footnote{The current Euro-VO partners are Astrogrid - UK, GAVO - the German Astrophysical Virtual Observatory, OV-France - the French Virtual Observatory, SVO - the Spanish Virtual Observatory, VObs.it - the Italian Virtual Observatory, working in close collaboration with ESA. Poland also has a national VO initiative (Stachowski et al., this issue \citep{Poland-VO}).} is a network of European national VO initiatives and international organizations. Each partner potentially brings different kinds of data centres with different management methods, and different technical expertise. Each national node and agency has its own organization, which depends on its priorities, which in turn depend on its funding sources, and on the way astronomy is organized in each country. In some countries, for instance, most VO resources are provided through a dedicated project; in others, there is a national coordination of the resources, which are provided by individual institutes and laboratories; in others, coordination is provided directly by a national funding agency. Many combinations of the above schemes are possible but the common characteristic is that all partners are strongly committed to VO development. The balance between the implementation of VO in data centres and technical work on the VO framework is also different for the different partners.

Euro-VO has proven to be indispensable for coordination and cooperation, and for the exchange of information and the sharing of expertise, strengthening the weight of the individual European projects in international cooperation. Individual projects keep their own aims, but learn from each other and build a global strategy, and duplication of efforts is minimized. This helps national activities to be better focused and to have much greater impact.

The European level also brings critical mass for the activities, in particular for support activities for scientists, data providers and education. The tutorials and support activities are developed at the European level, and then used as templates for national activities. Important collaborations between teams from different countries were also nurtured by the technological discussions within Euro-VO.

During the last few years, the European projects were funded to support solely coordination activities. The previous projects came also with  support for service activities and technological development. This was instrumental in focusing the activities in key elements of the VO, the essential initial standards and the tools which are now the core VO framework, besides the proper interests of the national and agency initiatives. In turn this empowered the European teams to influence the development of the global VO. There are currently 39 standards recommended by the IVOA, among which only two have no European co-author, and 22 have at least one European editor.

\section{Lessons learnt for best practices for VO activities}

As confirmed by the Euro-VO projects, the activities required to maintain the Virtual Observatory are the following:

\begin{itemize}
\item
Development of standards and protocols, and their international agreement;
\item
Construction of ``glue'' software components - registry, workflow, user authentication, virtual storage;
\item
Uptake by data centres, who need to ``publish'' to the system, i.e. to write VO compliant data services connected to their holdings;
\item
Construction of tools to effectively take advantage of seamless access to data, some of them able to act as VO portals for certain types of data;
\item
Support for the scientific community in its uptake of the new framework.
\end{itemize}

The three main components are well summarized in Figure 2, which shows the IVOA ``level zero'' architecture as stated in the IVOA Architecture document (Arviset et al. 2010 \citep{IVOAArchitecture}), with data providers (including modelling data) at the bottom and users (including machines) on the top, and in the box in the middle the elements of the Virtual Observatory framework: the VO provides a technical framework for the providers to share their data and services (``Sharing''), and allowing users to find (``Finding'') these resources, to get them (``Getting'') and to use them (``Using''). To enable these functionalities, the establishment of some core astronomically oriented standards (``VO Core'') is also necessary.

Euro-VO activities over the years led us to define best practices for the main strands of work: support to the scientific community, support to data providers, and coordination of technological work. They are detailed in the following sub-sections. Table 2 provides a summary of the activities organised by the VO projects described in the text : workshops, science calls, and the data centre census and its updates.

\begin{figure*}
\begin{center}
\includegraphics[width=10cm]{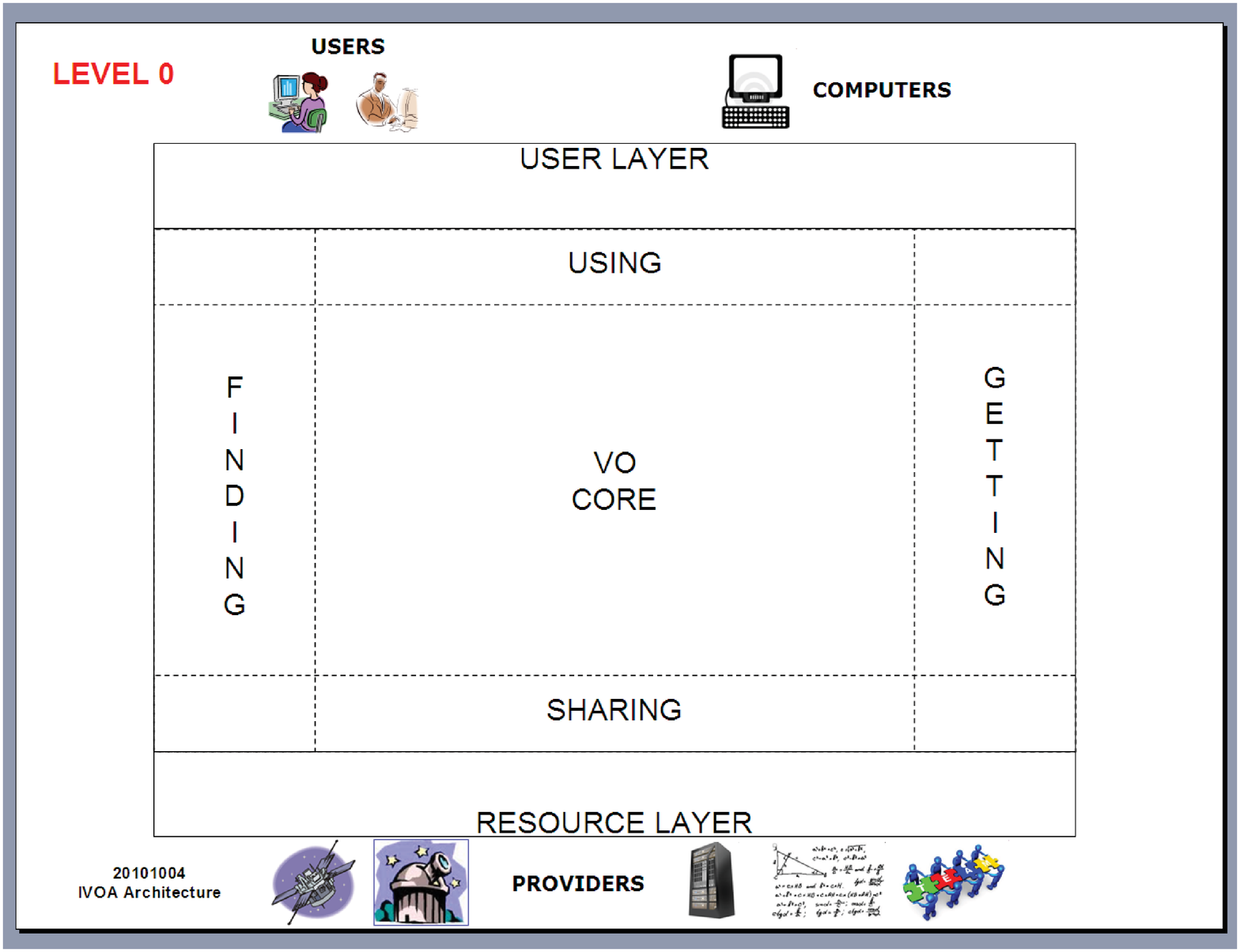}
\caption{IVOA ``level zero'' architecture (Arviset et al \citep{IVOAArchitecture})}
\end{center}
\end{figure*}

\begin{table}
\centering
\rotatebox{90}{
\begin{tabular}{c|c|p{10cm}|c}
Project & Date & Title (if relevant) & Location (if relevant) \\ \hline
\multicolumn{4}{l}{\emph{Topical scientific workshops}} \\ \hline
EuroVO-DCA & 2007 & Astronomical Spectroscopy and the Virtual Observatory & Villafranca (Spain) \\
 & 2008 & The Euro-VO Data Centre Alliance Theory Workshop & Garching (Germany) \\
 & 2008 & Euro-VO Data Centre Alliance workshop: Grid and Virtual Observatory & Garching (Germany) \\
EuroVO-AIDA & 2008 & Multi-wavelength astronomy and the Virtual Observatory & Villafranca (Spain) \\ \hline
\multicolumn{4}{l}{\emph{``Hands-on'' Workshops/VO Schools}} \\ \hline
EuroVO-AIDA & 2009 & & Garching (Germany) \\
 & 2010 & & Strasbourg (France) \\
EuroVO-ICE & 2011 & & Strasbourg (France) \\
CoSADIE & 2013 & & Villafranca (Spain) \\ \hline
\multicolumn{4}{l}{\emph{Data centre workshops}} \\ \hline
EuroVO-DCA & 2007 & How to publish data in the VO & Villafranca (Spain) \\
 & 2008 & How to publish data in the VO & Garching (Germany) \\
EuroVO-AIDA & 2009 & How to publish data in the VO & Villafranca (Spain) \\
CoSADIE & 2013 & CoSADIE Data Centre Forum & Heidelberg (Germany) \\ \hline
\multicolumn{4}{l}{\emph{Euro-VO Research Initiatives (Calls for scientific proposals}} \\ \hline
EuroVO-AIDA & 2008 & & \\
 & 2009 & & \\ \hline
\multicolumn{4}{l}{\emph{Data Centre Census and updates}} \\ \hline
EuroVO-DCA & 2008 & & \\
EuroVO-AIDA & 2010 & & \\
CoSADIE & 2013 & & \\ \hline
\end{tabular}
}

\caption{Euro-VO topical, hands-on/VO schools and data centre workshops, calls for scientific proposals, data centre census and updates, with the organising project and date, plus title and location if relevant.}
\end{table}

\subsection{Support to the scientific community}

Euro-VO considered from the beginning that providing support to the community includes gathering its requirements. The Science Advisory Committee was set up at a very early stage and regularly met to discuss science needs, comment on the Euro-VO activities and propose directions for future work. EuroVO-DCA also organised three topical workshops, \emph{Astronomical Spectroscopy and the Virtual Observatory} (2007, Villafranca - Guainazzi and Osuna 2008 \citep{2008asvo.proc.....G}) and two back-to-back meetings, the \emph{Theory} and \emph{Grid} Workshops (respectively Lemson et al. 2008 \citep{2009MmSAI..80..333L} and Taffoni et al. 2008 \citep{2009MmSAI..80..446T}), in Garching in April 2008. EuroVO-AIDA also organised one workshop, \emph{Multi-wavelength astronomy and the Virtual Observatory} (2008, Villafranca -  Baines and Osuna 2009 \citep{2009mavo.proc.....B}). Members of the science community were invited to discuss their scientific needs with VO scientists and developers, and  VO teams presented the current status of the VO in these domains. This was at a stage when the core standards and tools were being developed and it was important to gather detailed science needs.

For a more direct support to individual researchers, several kinds of activities were tested. The first projects included Calls for scientific proposals, the so-called \emph{Euro-VO Research Initiatives}, in which European teams were encouraged to submit proposals for research projects which could benefit from VO tools and applications and got support from VO teams to perform the research. This led to a few papers (e.g. Fathi el al. 2010 \citep{2010MNRAS.406.1595F}; Fathi 2010 \citep{2010ApJ...722L.120F}) and provided valuable experience on how to support scientific use of the VO.

While the Research Initiatives provided focused assistance to a few small groups in the European astronomical community, the most efficient way to reach the community and gather requirements has been ``Hands-On Workshops''. These were open to young astronomers from all over Europe, tutored by experienced VO scientists and developers. The workshops begin with tutorials which train participants to perform pre-defined science cases using VO-enabled data and tools. The participants are then invited to apply advanced usage of VO tools on their own research projects, and to present their results in the final plenary session. These Hands-On schools have proven to be very effective: the participants really learn how to use the VO for their own research subjects, and they share their knowledge when they are back in their labs; the tutors are able to gather feedback requirements in real time while observing very diverse usages of the data and tools. A set of tutorials\footnote{ \url{http://www.euro-vo.org/?q=science/scientific-tutorials}} has been maintained since 2007 and is regularly updated with new VO functionalities.

The template defined by the Euro-VO ``Hands-On Workshops'' has been used in national and regional events. There is a real interest in the scientific community, as demonstrated for instance by the remarkable success of the ``VO... in Tour'' organised by VObs.it, the Italian VO initiative, in 2010. VO Days were organised around the country in all INAF sites. These events gathered more than one-fourth of the Italian astronomers (244), and 176 of them filled in a feedback form (Pasian this issue \citep{VObs.it}). The Spanish VO SVO has also been organising high impact national VO schools\footnote{\url{ http://svo.cab.inta-csic.es/docs/index.php?pagename=Meetings}} since 2009. The eight schools gathered more than 200 participants (which represents more than one third of the census of Spanish astronomers). One year after the school, participants are contacted asking about the role VO-tools play in their daily research. In most of the cases  positive answers are received (they do really use VO-tools and services) and, in some cases, they have published VO-science papers in which they did the VO analysis independently (without support of SVO staff). European astronomers are currently playing a leading role in the VO science field.

\subsection{Support for data providers}

The EuroVO-DCA project performed a first Census of European data providers in 2008 (Allen et al. 2009 \citep{Census2009}), which showed a very diverse community, from large agencies in charge of ground and space-based telescopes, to small teams willing to share their knowledge by providing their results in the VO. The census was updated in 2010 and 2013. The initial census identified 69 Data Centres (updated to 79 in 2013) with a wide diversity in scale, content and function. The Data Centres were asked to identify the types of data and services they provide according to three main categories:  ``Observational Archives and Data Products'', ``Services, Tools and Software Suites'',  and ``Theory Archives or Services''. The Venn diagram in Figure 3 shows the relative proportions of the responses in each category, and the strong overlap between these categories with many Data Centres indicating that they provide multiple types of services. The Census is used to gather information on how data providers use the VO framework, and about their feedback and requirements.

\begin{figure*}
\includegraphics[width=7cm]{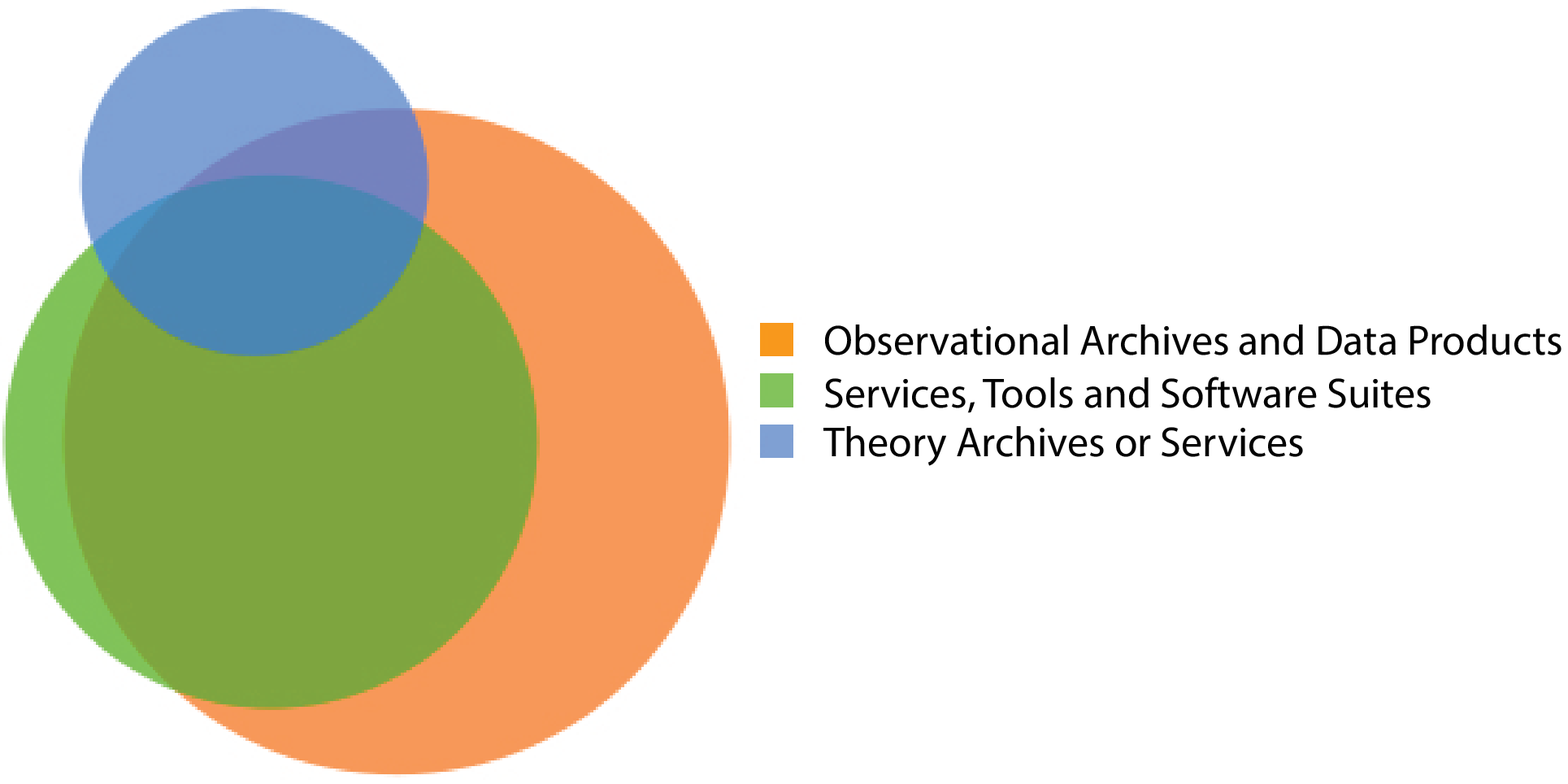}
\includegraphics[width=7cm]{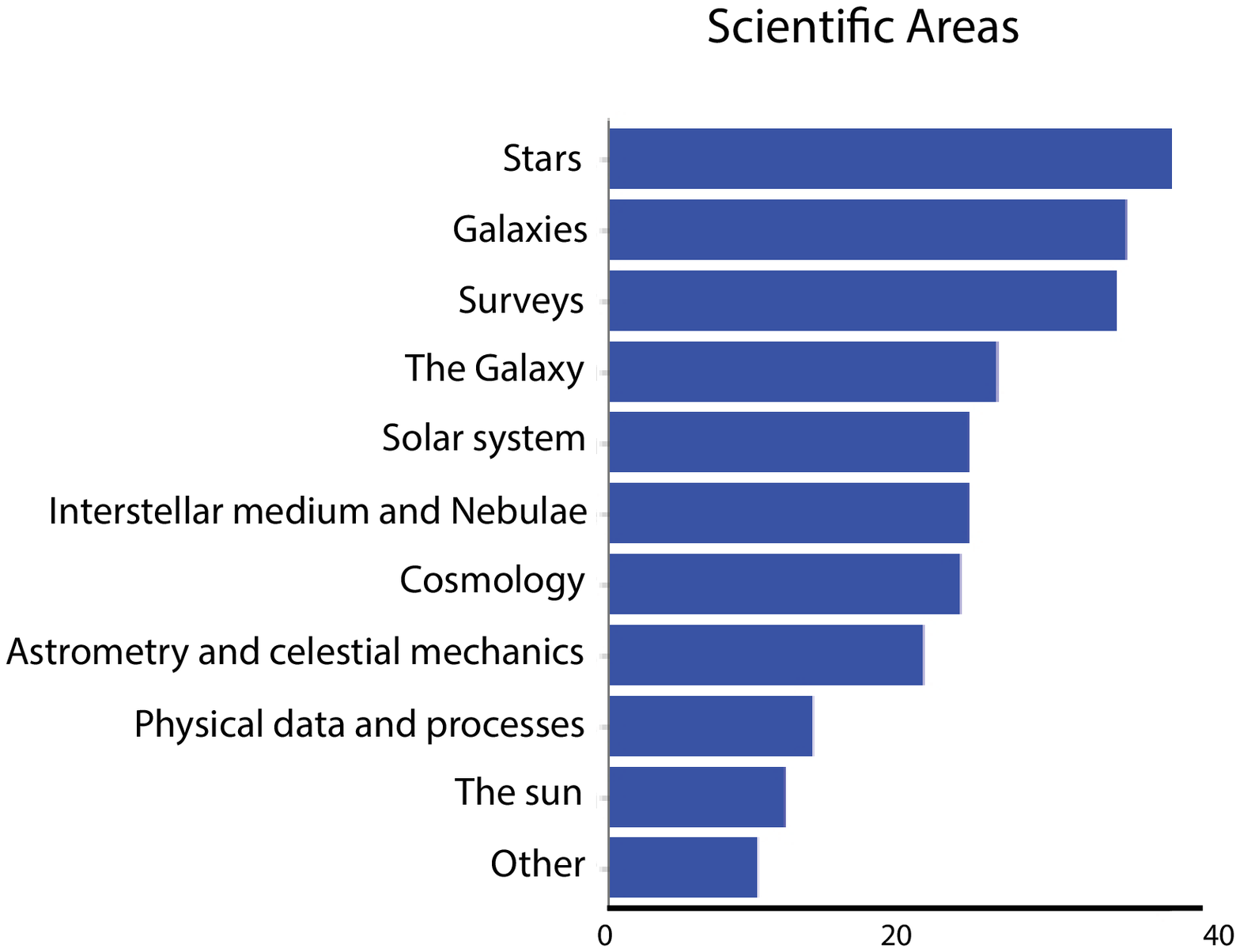}
\includegraphics[width=7cm]{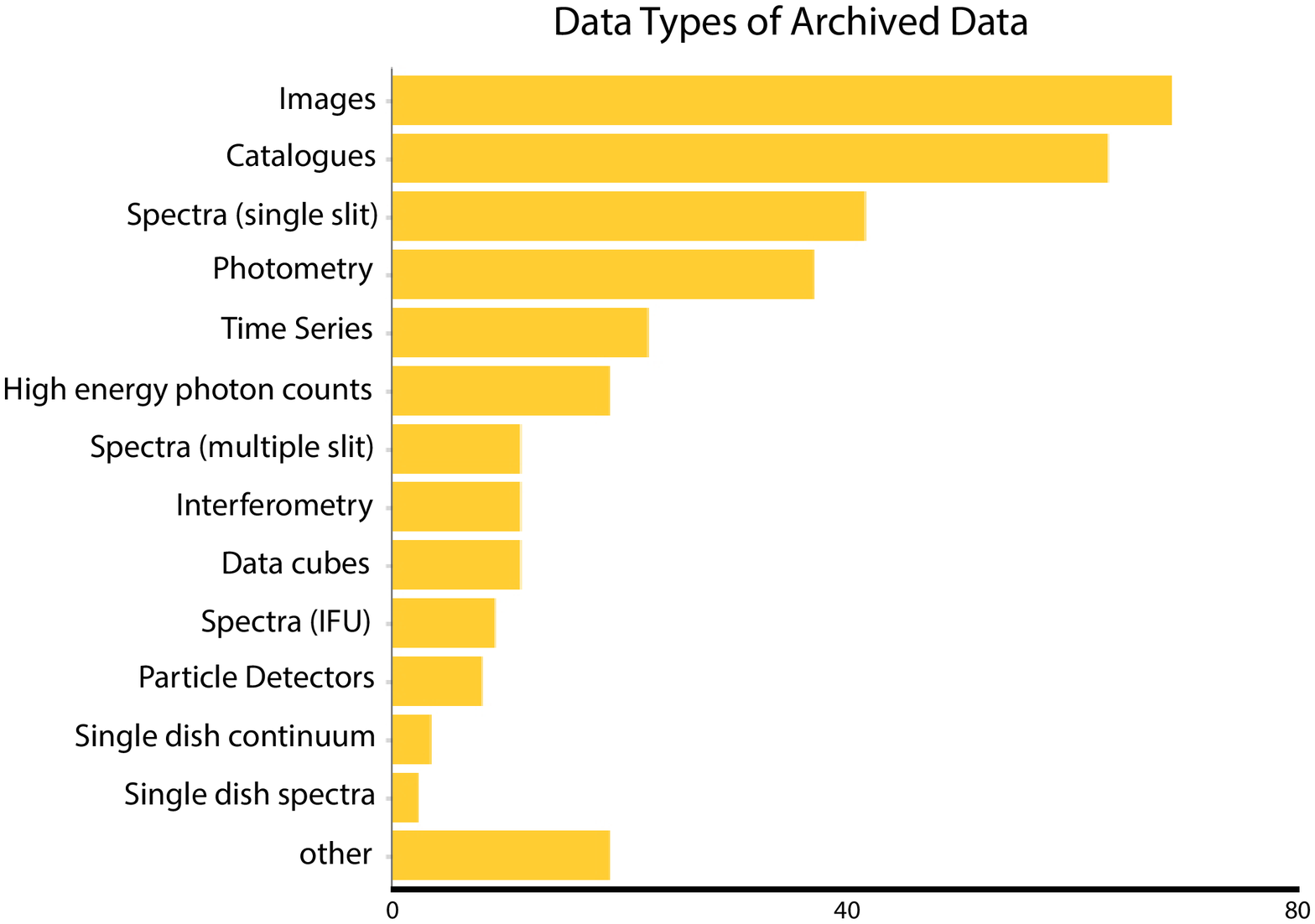}
\caption{(a) A Venn diagram of the DCA Census results on the main types of data and services provided by European Data Centres. This diagram represents some 69 Data Centres, where the areas of the three circles indicate the relative number of responses in each category.  The intersections of the circles indicate cases where multiple responses were received, showing that there is strong overlap in these different Data Centre activities. Of the 60 Data Centres who indicated that they provide 'Observational Archives and data Products', 35 also provide 'Services, Tools and Software Suites', and 10 also provide 'Theory Archives' or ‘Services’. (b) The Scientific Areas relevant to Data Centres reported in the census. (c) The types of data provided by Data Centres as reported in the census.
}
\end{figure*}

One main aim of the EuroVO-DCA project was to engage data providers, and two workshops on \emph{How to publish data in the VO} were organised in 2007 and 2008, with a similar event organised by EuroVO-AIDA in 2009. The workshops were open to participants from all over Europe. Data providers were trained in the usage of data publishing tools, and were invited to apply them to their own data holdings. Several participant groups and individuals eventually posted VO-compliant data and services, and some engaged as active IVOA participants. There was initially a good response to these workshops, but at the third workshop fewer participants applied and it was felt that this was the end of a cycle. The individual VO initiatives continued to give support to data providers on request, and many potential VO contributors also attended the IVOA meetings and events organised at the national level to learn about the VO framework.

The good level of dissemination of knowledge about the VO in the European data provider community was demonstrated by the \emph{Data Centre Forum} organised by the CoSADIE project, which gathered 40 participants in Heidelberg in June 2013. They were invited to explain how they use the VO framework, and many talks were presented, showing a high level of awareness and take-up. Data providers would welcome a regular forum to share their experience of using the VO, provide feedback and get support from the VO teams. It was emphasised that making data centre holdings available in the VO increases visibility and impact. Another important motivation for data providers to engage in VO take-up is that they are interested in having their data made useable in VO tools such as Aladin for visualization or TOPCAT for tabular data: they also use the VO because it saves time and effort and provides valuable tools. The help provided by VO teams was identified as an important factor for take-up, especially for the smaller data centres. Some also use VO publishing tools such as, currently, DACHS, SAADA, VODance, developed respectively in Germany, France and Italy, and the suite of tools developed by the Spanish VO (SVOCat, MySpec-MyImg)\footnote{Publishing tools have been developed since the early stages of the VO, for instance in the EuroVO-AIDA project, but most of the first generation of tools is obsolete.}. The most appropriate publishing tools depends on the characteristics, in particular of the complexity, of the data. VO publishing workshops are also organised at the national level, for instance by the French VO starting in 2004, by the Italian VO with a VO publishing workshop held in Trieste in February 2012, and by the Spanish VO in March 2014.

Large astronomy infrastructures also participated in these events, but the situation is a bit different for them. It is better and better understood that a close collaboration of the VO with the large projects is essential: they bring new kinds of data which may require specific developments and adaptations of the VO standards and tools. This has been identified as a key factor to ensure VO relevance and thus its sustainability in the medium term, and establishing links with these projects has been a target for the European VO projects. The IVOA has been organising Focus sessions with large current and future projects since its meeting in Heidelberg in 2013, and has set IVOA priorities with  the large recent and future projects in mind - these are currently multi-dimensional data (ALMA, VLT and E-ELT, LOFAR, SKA) and the time domain (LSST). European members of the IVOA Committee on Science Priorities have been among the driving forces of this evolution.

\subsection{Why and how to organize technical collaboration at a continental level}

The core coordination activity for the development of the VO framework of standards and tools is obviously at the IVOA level: the Working Groups and Intererest Groups, overseen by the Technical Coordination Group (TCG). IVOA organise the technical activities, and priorities are set up by the Executive Board. Also, as explained in Section 3 most of the technical work is performed by the national and agency VO initiatives.

The Euro-VO added-value for the technological activities is that it allowed the sharing of knowledge and the establishment of close collaboration in common technological activities, as well as the building of links between tools, which are one of the core assets of the VO interoperability framework thanks in particular to the SAMP\footnote{SAMP, the Simple Application Messaging Protocol, enables software tools to communicate and interoperate. SAMP supports communication between applications on the desktop and in web browsers, and is also intended to form a framework for more general messaging requirements.} standard (Taylor et al. 2012 \citep{2011arXiv1110.0528T}). As explained in the preamble of SAMP ``IVOA members have recognised that building a monolithic tool that attempts to fulfil all the requirements of all users is impractical, and it is a better use of our limited resources to enable individual tools to work together better''. Euro-VO fully endorsed this philosophy from the beginning and its partners put emphasis in the development of tools allowing access to, visualisation of and usage of VO-enabled data such as Aladin (VO portal for images), TOPCAT (Management tool for tabular data), VOSpec (visualisation and management of spectra), VOSA (Spectral energy distribution building and analysis).

Technical cooperation and coordination has been organised since the VOTECH project through bi-yearly meeting of the European technical teams involved in the VO development, the \emph{Euro-VO Technology Forums}. These meetings allowed the sharing of information about VO status and technical developments in the different countries, the establishment of cross-border collaborations, and also ``hands-on session'' (\emph{Hack-a-thon}) for testing or prototyping software. IVOA standards in development could also be discussed, especially when diverging opinions had shown up between European teams or when most of the interested participants were Europeans, allowing more efficient work in the IVOA context. Additional face-to-face meetings were also organised on specific topics, for instance theory - to which Euro-VO has been devoting special attention since the EuroVO-DCA project. These meetings were instrumental in building a Europe-wide technical VO community, including staff who were eventually not able to travel to the IVOA Interoperability meetings, in particular junior staff, temporary contractors and interns, who could get an early exposure to working in an international context.

The Euro-VO projects which provided support for technological activities also produced a detailed description of the VO architecture, and important  advances in the VO framework, by allowing the partners to focus technological work on key standards and tools and to assess new technologies.

\section{Euro-VO in a wider European context}

Euro-VO paid particular attention to liaisons with "nearby" disciplines. Several national initiatives, for instance the French and Italian ones include representatives of other disciplines, in particular planetary or heliospheric studies, or astroparticle physics, as well as representatives of the atomic and molecular physics (since atomic and molecular data are heavily used to interpret observations in several fields of astronomy). Several of these disciplines also set up European projects dealing with data aspects, for instance among others EuroPlanet, HELIO (and other projects assessing the sharing of heliospheric physics data) and VAMDC. The other disciplines assessed solutions developed for the astronomical Virtual Observatory with respect to their own needs. In some cases it did not fit but some of the VO standards and tools were re-used and customized in other disciplinary contexts, for instance standards such as the Registry of Resources, VOTable or the Table Access Protocol, or tools such as Aladin and TOPCAT.

The specific requirements of the Cerenkov Telescope Array CTA, a project common to the astronomy and astroparticle domains, are also tackled by Euro-VO and the IVOA to include these data fully in the Virtual Observatory context. They are for instance particularly useful to understand how to deal with ``provenance'' in the VO.

More generally, Euro-VO was one of the early examples of a European project dealing with the sharing of scientific data in a disciplinary context, and EuroVO-AIDA was one of the seven projects selected in the first Call for Proposals of the Seventh European Framework on the \emph{Scientific Digital Repositories} topic, in 2008. The astronomical VO has since then been presented in many meetings in Europe, as an example of an operational, world-wide disciplinary data framework. All disciplines have different cultures and organisations, and there is no unique solution to build the disciplinary part of the data infrastructure. But Euro-VO and the IVOA have explored many of the aspects of building a disciplinary framework and they share lessons learnt.

Liaison with projects which deal with the generic aspects of the data infrastructure was also sought by Euro-VO. Astronomy was one of the thematic topics of the first European project \footnote{This project (2012-2014), initially called iCORDI for \emph{International Collaboration on Research Data Infrastructure}, soon became RDA/Europe.} devoted to the European support for the emerging Research Data Alliance RDA. The aims were both to share lessons learnt in the building of the IVOA, and to explore possible liaison with generic interoperability projects, in particular EUDAT\footnote{\url{www.eudat.eu}} (European project which builds generic building blocks for the scientific data infrastructure). These were also among CoSADIE goals.

The Virtual Observatory framework has never been an isolated island, and generic building blocks were used whenever possible. In particular, the IVOA Registry of Resources (Hanisch et al., 2007, \citep{2011arXiv1110.0514H}) is compliant with OAI-PMH, the \emph{Open Archive Initiative - Protocol for Metadata Harvesting} widely used by the digital library community (and also by EUDAT). Also the standard format for IVOA vocabularies (Derriere et al., 2009 \citep{2011arXiv1110.0520D}) is based on the W3C's Resource Description Framework (RDF) and Simple Knowledge Organization System (SKOS). As stated in the IVOA Recommendation for vocabularies, ``by adopting a standard and simple format, the IVOA will permit different groups to create and maintain their own specialised vocabularies while letting the rest of the astronomical community access, use, and combine them. The use of current, open standards ensures that VO applications will be able to tap into resources of the growing semantic web.'' The aim of Euro-VO liaison with EUDAT and the RDA is also to make sure that the generic framework is able to interoperate with existing disciplinary frameworks, and that our specific requirements are taken into account.

\section{Conclusion}

As shown throughout this paper, the three main pillars to build and maintain VO coordination are support for scientific users, support for data providers, and coordination of technical activities. At the beginning of the Euro-VO endeavour, this was described by a proposed organisation with three entities, the \emph{Euro-VO Facility Centre} (in support of the scientific users), the \emph{Euro-VO Data Centre Alliance} and the \emph{Euro-VO Technology Centre}, each led by one or two pre-defined partners, with distributed participation (Padovani, 2006 \citep{2006ASPC..351..771P}; Genova, 2009 \citep{2009MmSAI..80..548G}). Euro-VO partnership evolved along the years, and the current system is more versatile, with a Memorandum of Understanding between five organisations and institutes currently representing the national initiatives which share the core VO activities in Europe\footnote{CNRS/INSU, INAF, INTA, the Universities of Edinburgh and Heidelberg} in close liaison with ESA, and possibly evolving responsibilities for each partner defined in common agreement.

The Euro-VO project has been linking together the European national and agency VO initiatives and increasing their effectiveness. It succeeded in building a European-wide VO community of scientific users, data providers and VO teams, in making astronomy visible amongst the European disciplinary data infrastructures, and in having the Virtual Observatory recognized as a Research Infrastructures of astronomy in the European Roadmap established by the Astronet ERA-NET in 2008 \cite{Roadmap2008}. It has proven to be indispensable for cooperation and collaboration.

One of its products is the assessment of all the elements needed for medium term sustainability, and work is on-going, in particular with the funding agencies represented in Astronet, to understand how these elements can be put in place. One clear evolution is the need to establish closer links with the large projects of the discipline. This is the aim of the next Euro-VO project, which enters as a Work Package, WP4 \emph{Data Access, Discovery and Interoperability} of the ASTERICS project, currently (February 2015) in negociation with the European Commission for a start on May 1st, 2015 and a 4 year duration. ASTERICS (Astronomy ESFRI\footnote{ESFRI is the European Strategic Forum for Research Infrastructures, which develops a research infrastructure roadmap which includes all scientific fields. The 2010 Roadmap for astronomy includes E-ELT, CTA, SKA and KM3Net.}\& Research Infrastructure Cluster) ``aims to address the cross-cutting synergies and common challenges shared by the Astronomy and Astroparticle ESFRI facilities''. ASTERICS WP4 will establish a close collaboration between the Euro-VO and infrastructure teams. It will develop seamless access to the data products of the ESFRI projects and pathfinders, and other large infrastructures such as the European Gravitational Observatory EGO, in the VO framework. The tasks again cover support to data providers, support to scientific usage, and technological development.

Another clear strand for evolution is a closer collaboration with, and effective participation in, projects which work on the generic framework for data sharing, driven in particular by the very fast emergence of the Research Data Alliance as a major global initiative in scientific data sharing.

\section{Acknowledgements}

The Euro-VO project has received funding from the European Union Fifth, Sixth and Seventh Framework Programmes for research, technological development and demonstration under grant agreements no HPRI-CT-1999-4002, HPRI-CT-2001-50030 (FP5), 011892, RI031675 (FP6), 212104, 261541, 312559 (FP7).




\section*{References}

\bibliography{mybibfile}

\begin{thebibliography}{10}
\bibitem{VObs.it}
F.~Pasian, Vobs.it, the italian contribution to the international virtual
  observatory -- history, activities, strategy, Astronomy and Computing this
  issue.
\end{thebibliography}

\end{document}